\documentclass{elsart}
\usepackage{amssymb}
\usepackage{amsmath}

\newtheorem{definition}{Definition}[section]
\newtheorem{lemma}[definition]{Lemma}
\newtheorem{theorem}[definition]{Theorem}
\newtheorem{proposition}[definition]{Proposition}
\newtheorem{corollary}[definition]{Corollary}
\newtheorem{remark}[definition]{Remark}
\newtheorem{example}[definition]{Example}

 1   
\font\ddpp=msbm10  scaled \magstep 1  

\def\QED{\hskip0.1em\hfill\null\ \null\nobreak\hfill
\kern3pt\lower1.8pt\vbox{\hrule\hbox   {\vrule\kern1pt\vbox{\kern1.7pt
\hbox{$\scriptstyle   QED$}\kern0.2pt}\kern1pt\vrule}\hrule}}
\def\R{\hbox{\ddpp R}}               
\def\L{\hbox{\ddpp L}}

\def\theequation{\arabic{section}.\arabic{equation}}
\def\blob{\quad\rule{8pt}{8pt}}

\begin{document}

\begin{frontmatter}

\title{Tulczyjew's triples and lagrangian submanifolds in classical field theories}

\author{Manuel de Le\'on, David Mart{\'\i}n de Diego,}
\author{Aitor Santamar{\'\i}a-Merino}

\address{Instituto de Matem\'aticas y F\'\i sica Fundamental\\
Consejo Superior de Investigaci\'on Cient\'\i fica\\
Serrano 123, 28006 Madrid, Spain\\
email: mdeleon@imaff.cfmac.csic.es,
d.martin@imaff.cfmac.csic.es\\ 
aitors@imaff.cfmac.csic.es}

\begin{abstract}
In this paper the notion of Tulczyjew's triples in classical mechanics
is extended to classical field theories, using the so-called multisymplectic formalism,
and a convenient notion of lagrangian submanifold in multisymplectic geometry.
Accordingly, the dynamical equations are interpreted as the local equations
defining these lagrangian submanifolds.
\end{abstract}
\end{frontmatter}

\section{Introduction}

In middle seventies, W.M. Tulczyjew \cite{T1,T2} introduced
the notion of special symplectic manifold, which is a symplectic manifold symplectomorphic
to a cotangent bundle. Using this notion, Tulczyjew gave a nice interpretation of
lagrangian and hamiltonian dynamics as lagrangian submanifolds of convenient
special symplectic manifolds.

The other ingredients in the theory were two canonical diffeomorphisms
$\alpha : TT^*Q \longrightarrow T^*TQ$ and $\beta: TT^*Q \longrightarrow T^*T^*Q$.
$\beta$ is nothing but the mapping obtained by contraction with
the canonical symplectic form $\omega_Q$, but the definition of $\alpha$ is
more complicated, and requires the use of the canonical involution of the
double tangent bundle $TTQ$.

The theory was extended to higher order mechanics by several authors
(see for instance \cite{CCS,CCSS,CLL,crampin,LL}). 
But the extension to classical field theories has not been achieved up
to now. There is a good approach by Kijowski and Tulczyjew \cite{KT},
and in fact, the present approach is strongly inspired in that monograph.

The key point is a better understanding of the geometry of lagrangian submanifolds
in the multisymplectic setting. A systematic study of the geometry of multisymplectic
manifolds was started by Cantrijn {\sl et al} at the beginning of the nineties \cite{CCI},
followed by a pair of papers which clarify that geometry \cite{CIL0,CIL1}
(a more detailed study \cite{aitor2} is in preparation).

A multisymplectic manifold is a manifold equipped with a closed form
which is non-degenerate in some sense. The canonical examples are the bundles 
of forms on an arbitrary manifold, providing thus
a nice extension of the notion of symplectic manifold.
However, this definition is too
general for practical purposes. Indeed, in order to have a Darboux theorem
which would permit us to introduce canonical coordinates, we need additional properties.
In other words, if we want to deal with multisymplectic manifolds
which locally behave as the geometric models we need to
consider multisymplectic manifolds $({\mathcal P}, \Omega)$ with additional structure, given by
a 1-isotropic foliation ${\mathcal W}$ satisfying some
dimensionality condition, or, even a ``generalised foliation"
${\mathcal E}$ defined roughly speaking on the space of leaves determined by ${\mathcal W}$.

The tangent and cotangent functors are now substituted by
the jet prolongation functor and the exterior power functor, respectively, so
that we obtain canonical diffeomorphisms
$\tilde{\alpha} : \widetilde{J^1Z^*} \longrightarrow \Lambda^{n+1}_2Z$
and
$\tilde{\beta} : \widetilde{J^1Z^*} \longrightarrow \Lambda^{n+1}_2Z^*$,
where $Z$ is the 1-jet prolongation of the fibred manifold
$Y \longrightarrow X$, $X$ being the space-time $n$-dimensional manifold, and $Z^*$ is
the dual affine bundle of $Z$. Here a tilde over a manifold of jets
means that we are taking a quotient manifold in order
to have only those elements with the same divergence.

Using a convenient formulation of the field equations
with Ehresmann connections, we construct the corresponding 
lagrangian submanifolds which encode the dynamics.
Indeed, we present a compact form for the De Donder and field equations
as follows. From the lagrangian density $\L = L\eta$ ($\eta$ is a volume form on $X$),
we construct the Poincar\'e-Cartan $(n+1)$-form $\Omega_L$ 
on $Z$; then the extremals for $\L$ coincide with the horizontal 
sections of any Ehresmann connection ${\bf h}$ in the fibred manifold
$Z \longrightarrow X$ satisfying the equation
$$
i_{\bf h} \, \Omega_L = (n-1) \Omega_L.
$$
Since a connection in $Z \longrightarrow X$ can be interpreted as a
section of the 1-jet prolongation $J^1Z \longrightarrow Z$, we have
all the ingredients we need. In fact, the Euler-Lagrange equations
are just the local equations defined by a $k$-lagrangian submanifold
of $\widetilde{J^1Z^*}$, the latter being a multisymplectic manifold
equipped with the multisymplectic form
$\Omega_\alpha$ dragged via
$\tilde{\alpha}$ from the canonical one on $\Lambda^{n+1}_2Z$. 

A similar procedure can be developed in the hamiltonian setting, but
in this case we would need to choose a convenient hamiltonian form. 
This hamiltonian form is obtained through the corresponding Legendre 
transformation $Leg_L : Z \longrightarrow Z^*$.
Finally, both sides are related.

\section{Lagrangian submanifolds and classical mechanics}

\subsection{Some prelimaries}

Let $({\mathcal V}, \omega)$ a finite
dimensional symplectic vector space with symplectic form $\omega$.
This means that $\omega$ is a 2-form on a vector space
$V$ which is non-degenerate in the sense that
the linear mapping
$$
v \in {\mathcal V} \mapsto i_v \, \omega \in V^*
$$
is injective (and hence it is a linear isomorphism). 

Therefore, ${\mathcal V}$ has even dimension, say $2n$, and the non-degeneracy
is equivalent to the condition $\omega^n \not= 0$.

A linear isomorphism $\phi : ({\mathcal V}_1, \omega_1) \longrightarrow ({\mathcal V}_2, \omega_2)$
is called a symplectomorphism if $\phi$ preserves the symplectic forms,
say $\phi^* \omega_2 = \omega_1$.

Take a subspace $E \subset {\mathcal V}$, and define the $\omega$-complement of $E$ as
follows:
$$
E^\perp = \{v \in {\mathcal V} \; | \; i_{v \wedge e} \,  \omega = 0, \;
\hbox{for all} \; e \in E \}.
$$ 

The subspace $E$ is said to be isotropic (resp. coisotropic, lagrangian, symplectic)
if $E \subset E^\perp$ (resp.
$E^\perp \subset E$, $E = E^\perp$, $E \cap E^\perp = \{0\}$).

An useful characterization of a lagrangian subspace $E$,
is that it is a maximally isotropic subspace or, equivalently, 
on a finite dimensional symplectic vector space, it is isotropic and 
$\displaystyle{\dim E = \frac{1}{2} \dim {\mathcal V}}$.

The algebraic model for a symplectic vector space is the following. Given an arbitrary
vector space $V$ we construct ${\mathcal V}_V = V \oplus V^*$ equipped
with the symplectic form $\omega_V$ defined by
$$
\omega_V((v_1, \gamma_1), (v_2, \gamma_2)) = \gamma_1(v_2) - \gamma_2(v_1),
$$
for all $(v_1, \gamma_1), (v_2, \gamma_2) \in {\mathcal V}_V$.

We know that $V$ and $V^*$ are lagrangian subspaces of $({\mathcal V}_V, \omega_V)$.
Moreover, every symplectic vector space $({\mathcal V}, \omega)$ is symplectomorphic
to $({\mathcal V}_{\mathcal L}, \omega_{\mathcal L})$ for any lagrangian subspace
${\mathcal L}$ of $({\mathcal V}, \omega)$.

In addition we can prove that a linear isomorphism
$\phi : ({\mathcal V}_1, \omega_1) \longrightarrow ({\mathcal V}_2, \omega_2)$
is a symplectomorphism if and only if its graph 
$\{(v, \phi(v)) \, |\, v \in {\mathcal V}_1\} \subset {\mathcal V}_1 \times {\mathcal V}_2$
is a lagrangian subspace
of the symplectic manifold
$({\mathcal V}_1 \times {\mathcal V}_2, \omega_1 \ominus \omega_2)$,
where $\omega_1 \ominus \omega_2 = \pi_1^*\omega_1 - \pi_2^* \omega_2$,
$\pi_1 : {\mathcal V}_1, \times {\mathcal V}_2 \longrightarrow
{\mathcal V}_1$ and $\pi_2 : {\mathcal V}_1, \times {\mathcal V}_2 \longrightarrow
{\mathcal V}_2$ being the canonical projections.

\medskip

A symplectic manifold is a pair $({\mathcal P}, \omega)$, where $\omega$ is 
a closed 2-form such that the pair $(T_x{\mathcal P}, \omega_x)$ is
a symplectic vector space for any $x \in {\mathcal P}$. Thus, ${\mathcal P}$ 
has even dimension, say $2n$.

Therefore, given a function $f : {\mathcal P} \longrightarrow \R$ there exists
a unique vector field (the hamiltonian vector field $X_f$ with
hamiltonian energy $f$) such that
$$
i_{X_f} \, \omega = df.
$$

Let now $\pi_Q : T^*Q \longrightarrow Q$ be the cotangent bundle of 
an arbitrary manifold $Q$. There exists a canonical 1-form $\theta_Q$ on $T^*Q$
defined by
$$
\theta_Q(\gamma)(X) = \langle \gamma, T\pi_Q(X) \rangle
$$
for all $X \in T_{\gamma}(T^*Q)$ and for all $\gamma \in T^*Q$.
$\theta_Q$ is the Liouville 1-form, and in bundle coordinates
$(q, p)$ we have
$$
\theta_Q = p dq.
$$

So, $\omega_Q = - d\theta_Q$ is a canonical symplectic form on $T^*Q$
such that $\omega_Q = dq \wedge dp$.

As is well known, Darboux theorem states that any symplectic manifold
is locally symplectomorphic to a cotangent bundle. More precisely,
one can find local coordinates around each point of a symplectic manifold
$({\mathcal P}, \omega)$ such that
$$
\omega = dq \wedge dp.
$$

The following results are the main examples of lagrangian submanifolds.

\begin{theorem}
$\,$
\begin{enumerate}
\item The image of a hamiltonian vector field $X_f$ on a symplectic
manifold $({\mathcal P}, \omega)$ is a lagrangian submanifold of the tangent 
lift symplectic manifold $(T{\mathcal P}, \omega^T)$.
\item The fibres of $T^*Q$ are lagrangian submanifolds of $(T^*Q, \omega_Q)$.
\item The image of a 1-form $\gamma$ on a manifold $Q$ is a lagrangian submanifold
of $(T^*Q, \omega_Q)$ if and only if $\gamma$ is closed.
\item Given a diffeomorphism 
$\phi : ({\mathcal P}_1, \omega_1) \longrightarrow ({\mathcal P}_2, \omega_2)$
between two symplectic manifolds then $\phi$ is a symplectomorphism
if and only if its graph is a lagrangian submanifold in
the symplectic manifold $({\mathcal P}_1 \times {\mathcal P}_2, \omega_1 \ominus \omega_2)$. 
\end{enumerate}
\end{theorem}

There is an important theorem due to A. Weinstein which gives 
the normal form for a lagrangian submanifold ${\mathcal L}$ in a 
symplectic manifold $({\mathcal P}, \omega)$.

\begin{theorem}\label{teorema-3}
Let $({\mathcal P}, \omega)$ be a symplectic manifold, and
let ${\mathcal L}$ be a lagrangian
submanifold. Then there exists
a tubular neighbourhod $U$ of ${\mathcal L}$ in ${\mathcal P}$, and a diffeomorphism
$\phi : U \longrightarrow V=\phi(U) \subset  T^* {\mathcal L}$ into
an open neighbourhood $V$ of the zero cross-section in $T^*{\mathcal L}$
such that $\phi^* \omega_{\mathcal L} = \omega_{|U}$, where
$\omega_{\mathcal L}$ is the canonical symplectic form on
$T^*{\mathcal L}$.
\end{theorem}

\subsection{Lagrangian and hamiltonian dynamics}

We shall recall the main results, more details can be found in \cite{LR}.

Let $L : TQ \longrightarrow \R$ be a lagrangian function.
We construct a 2-form $\omega_L$ by putting
$$
\omega_L = - d\theta_L
$$
where $\theta_L = S^*(dL)$. Here $S^*$ is the adjoint operator
of the canonical vertical endomorphism 
$\displaystyle{S = dq \otimes \frac{\partial}{\partial \dot{q}}} \,$.
We have omitted the indices of the coordinates, and used the notation
$(q, \dot{q})$ for the bundle coordinates on the tangent bundle
$\tau_Q : TQ \longrightarrow Q$.

The energy function is defined by
$$
E_L = \Delta(L) - L
$$
where $\displaystyle{\Delta = \dot{q} \frac{\partial}{\partial \dot{q}}}$ is
the Liouville or dilation vector field.

In local coordinates we have
$$
\omega_{L} = dq \wedge d\hat{p}, \qquad E_{\mathcal L} = \dot{q} \hat{p} - L,
$$
where $\displaystyle{\hat{p} = \frac{\partial L}{\partial \dot{q}}}$.
The lagrangian is regular if and only if the hessian matrix
$$
\left(
\frac{\partial^2 L}{\partial \dot{q}^i \partial \dot{q}^j} 
\right)
$$
is non-singular, where $i, j =1, \dots, n= \dim \, Q$.

We have that $L$ is regular if and only if $\omega_{L}$ is symplectic.
In such case, there is a unique vector field $\xi_{L}$ satisfying the equation
\begin{equation}\label{eq1}
i_{\xi_{L}} \, \omega_{L} = dE_{L}.
\end{equation}

$\xi_{L}$ is a second order differential equation on $TQ$ such that its
solutions (the curves in $Q$ whose lifts to $TQ$ are
integral curves of $\xi_{L}$) are just the solutions
of the Euler-Lagrange equations for ${L}$:
\begin{equation}\label{eq2}
\frac{d}{dt}\left(\frac{\partial {L}}{\partial \dot{q}} \right)
- \frac{\partial {L}}{\partial q} = 0.
\end{equation}

\medskip

Let now $H : T^*Q \longrightarrow \R$ be a hamiltonian function.
We denote by $X_H$ the corresponding hamiltonian vector field with respect
to $\omega_Q$.
In bundle coordinates we have
$$
X_H = \frac{\partial H}{\partial p} \frac{\partial}{\partial q}
- \frac{\partial H}{\partial q} \frac{\partial}{\partial p}
$$
Therefore, the integral curves $(q(t), p(t))$ of $X_H$ satisfy
the Hamilton equations
\begin{eqnarray*}
  \frac{dq}{dt} & = & \frac{\partial H}{\partial p} \\
 \frac{dp}{dt} & = & - \frac{\partial H}{\partial q}
\end{eqnarray*}

\medskip

The lagrangian and hamiltonian formalisms are connected through
the Legendre transformation. More precisely,
given a lagrangian function 
${L} : TQ \longrightarrow \R$
we define a fibred mapping $Leg_{L} : TQ \longrightarrow T^*Q$ over $Q$
by
$$
Leg_{L} (q, \dot{q}) = (q, \frac{\partial {L}}{\partial \dot{q}}).
$$
We know that ${L}$ is regular if and only if $Leg_{L}$ is 
a local diffeomorphism. For simplicity, we will assume that
${\mathcal L}$ is hyperregular, which means that
$Leg_{L}$ is a diffeomorphism.
In such case, $Leg_{L}$ is in fact a symplectomorphism
and, therefore, $\xi_{L}$ and $X_H$ are $Leg_{L}$-related,
when $H = E_{L} \circ {Leg_{L}}^{-1}$.
As a consequence, the Euler-Lagrange equations are translated
into the Hamilton equations via $Leg_{L}$.

\subsection{Dynamics as lagrangian submanifolds}

In \cite{T1,T2} W.M. Tulczyjew defined two canonical diffeomorphisms
\begin{eqnarray*}
\alpha &:& TT^*Q \longrightarrow T^*TQ\\
\beta &:& TT^*Q \longrightarrow T^*T^*Q
\end{eqnarray*}
locally given by
\begin{eqnarray*}
\alpha (q, p, \dot{q}, \dot{p}) &=& (q, \dot{q}, \dot{p}, p)\\
\beta  (q, p, \dot{q}, \dot{p}) &=& (q, p, -\dot{p}, \dot{q})
\end{eqnarray*}
with the obvious notations, where we have omitted the indices
for the sake of simplicity.

The second diffeomorphism is nothing but the contraction
with the canonical symplectic form $\omega_Q$ on $T^*Q$. The intrinsic definition
of $\alpha$ is more involved, and we remit to \cite{T1} for details.
We have the following commutative
diagram which justifies the name of Tulczyjew' s triple
for the above construction:

\begin{figure}[h]
\centering
\setlength{\unitlength}{1cm}
\begin{center}
\begin{picture}(4,2.5)(-0.7,0)
\put(-4,2){\makebox(0,0)[r]{$T^*TQ$}}
\put(0.3,2){\makebox(0,0)[r]{$TT^*Q$}}
\put(4,2){\makebox(0,0)[l]{$T^*T^*Q$}}
\put(2,0){\makebox(0,0)[c]{$T^*Q$}}
\put(-2,0){\makebox(0,0)[c]{$TQ$}}

\put(0.4,2){\vector(1,0){3.4}}
\put(-1,2){\vector(-1,0){2.8}}

\put(2,2.3){\makebox(0,0)[r]{$\beta$}}
\put(-2.5,2.3){\makebox(0,0)[l]{$\alpha$}}

\put(0.2,1.7){\vector(1,-1){1.4}}
\put(3.8,1.7){\vector(-1,-1){1.4}}
\put(-3.8,1.7){\vector(1,-1){1.4}}
\put(-0.2,1.7){\vector(-1,-1){1.4}}

\put(0.9,0.8){\makebox(0,0)[r]{$\tau_{T^*Q}$}}
\put(3.2,0.8){\makebox(0,0)[l]{$\pi_{T^*Q}$}}
\put(-1.3,1){\makebox(0,0)[r]{$T\pi_Q$}}
\put(-4,0.8){\makebox(0,0)[l]{$\pi_{TQ}$}}

\end{picture}
\end{center}
\centering
\end{figure}

The manifold $TT^*Q$ is endowed with two
symplectic structures, in principle different. Indeed, they are
$\omega_\alpha=\alpha^* \omega_{TQ}$ and $\omega_\beta=\beta^*\omega_{T^*Q}$.
A direct computation shows that both coincide up to the sign
(say $\omega_\alpha + \omega_\beta =0$), and, in addition, that
the symplectic form $\omega_\alpha$ is nothing but the complete 
or tangent lift $\omega_Q^T$
of $\omega_Q$ to $TT^*Q$.

We denote by
$\theta_\alpha = \alpha^*\theta_{TQ}$
and 
$\theta_\beta = \beta^*\theta_{T^*Q}$, such that
$\omega_\alpha = - d\theta_\alpha$ and
$\omega_\beta = - d \theta_\beta$.
In local coordinates we have
\begin{eqnarray*}
\theta_\alpha &=& \dot{p} dq + pd\dot{q}
\\
\theta_\beta &=& - \dot{p}dq + \dot{q} dp
\end{eqnarray*}

In fact, $TT^*Q$, equipped with the symplectic form $\omega_\alpha=-\omega_\beta=\omega_Q^T$
is an example of special symplectic manifold according to
the definition introduced by Tulczyjew in \cite{T1}.

\begin{definition}\label{especial0}
A {\rm special symplectic manifold} is a symplectic manifold
$({\mathcal P}, \omega)$ which is symplectomorphic to a cotangent bundle.
More precisely, there exists a fibration $\pi : {\mathcal P} \longrightarrow M$, and a 1-form $\theta$ on ${\mathcal P}$, such that
$\omega=-d\theta$, and $\alpha : {\mathcal P} \longrightarrow T^*M$ is a 
diffeomorphism such that $\pi_M \circ \alpha = \pi$ and $\alpha^*\theta_M = \theta$.
\end{definition}

The following is an important result for our discussion.

\begin{theorem}\label{subvala}
Let $({\mathcal P}, \omega=-d\theta)$ an special symplectic manifold, 
let $f: M \longrightarrow \R$ be a function, and denote by
$N_f$ the submanifold of ${\mathcal P}$ where $df$ and $\theta$ coincide.
Then $N_f$ is a lagrangian submanifold of $({\mathcal P}, \omega)$
and $f$ is a generating function.
\end{theorem}

Theorem \ref{subvala} applies to the particular case of Mechanics.
Indeed, if we consider a lagrangian function $L : TQ \longrightarrow \R$
we obtain a lagrangian submanifold $N_L$ of the symplectic
manifold $(TT^*Q, \omega_\alpha)$ with generating function $L$.

\medskip

Now, assume that $H : T^*Q \longrightarrow \R$ is a hamiltonian function,
with hamiltonian vector field $X_H$.

We have the following results.

\begin{theorem}
$\,$
\begin{enumerate}
\item The image of $X_H$ is a lagrangian submanifold of
$(TT^*Q, \omega_\alpha)$.
\item The image of $dH$ is a lagrangian submanifold of $(T^*T^*Q, \omega_{T^*Q})$.
\item $\beta(\hbox{Im} \;  X_H) = \hbox{Im} \; dH$.
\end{enumerate}
\end{theorem}

Finally, we relate both lagrangian submanifolds $N_L$ and $\hbox{Im} \; X_H$.

\begin{theorem}
Let $H$ be the hamiltonian function corresponding
to the hyperregular lagrangian function $L$, say $H = E_L \circ Leg_L^{-1}$.
Then we have
$N_L = \hbox{Im} \;  X_H$.
\end{theorem}

\section{Multisymplectic manifolds and their lagrangian sub\-mani\-folds}

\subsection{Multisymplectic vector spaces}

\begin{definition}
Let $\Omega$ be a $(k+1)$-form on a vector pace ${\mathcal V}$.
The pair $({\mathcal V}, \Omega)$ is called a {\rm multisymplectic vector space}
if the form $\Omega$ is non-degenerate, that is, the linear mapping
$$
v \in {\mathcal V} \mapsto i_v \Omega \in \Lambda^k {\mathcal V}^*
$$
is injective. The form $\Omega$ is called {\rm multisymplectic}.
\end{definition}

Let $({\mathcal V}_1, \Omega_1)$ and
$({\mathcal V}_2, \Omega_2)$ be two multisymplectic vector spaces  (of the same order $(k+1)$)
and 
let $\phi : ({\mathcal V}_1, \Omega_1) \longrightarrow ({\mathcal V}_2, \Omega_2)$ be
a linear isomorphism.

\begin{definition}
$\phi$ is called a {\rm multisymplectomorphism} if it preserves the
multisymplectic forms, i.e.
$\phi^* \Omega_2 = \Omega_1$.
\end{definition}

\begin{example}{\rm
Let $V$ be an arbitrary vector space and consider the direct product
${\mathcal V}_V = V \times \Lambda^kV^*$. Define a $k$-form $\Omega_V$ on ${\mathcal V}_V$ as
follows:
$$
\Omega_V ((v_1, \gamma_1), \dots, (v_{k+1}, \gamma_{k+1}))
= \sum_{i=1}^{k} \, (-1)^i \gamma_i(v_1, \dots, \check{v}_i, \dots, v_{k+1}),
$$
for all $(v_i, \gamma_i) \in {\mathcal V}_V$, $i=1, \dots, k+1$,
where a check accent over a letter means that it is omitted.
A direct computation shows that $\Omega_V$ is indeed multisymplectic.

If $E$ is a vector subspace of $V$, we consider the subspace
${\mathcal V}_V^r = V \times \Lambda^k_rV^*$, where $\Lambda^k_rV^*$
denotes the space of $k$-forms on $V$ vanishing when applied to at least
$r$ of their arguments from $E$.
Of course, ${\mathcal V}_V^r$ equipped with the
restriction $\Omega^r_V$ of $\Omega_V$ to ${\mathcal V}_V^r$
is a multisymplectic vector space.
If $E=\{0\}$ we recover ${\mathcal V}_V$.
}
\end{example}

Let $({\mathcal V}, \Omega)$ be a multisymplectic vector space of order $k+1$, and
${\mathcal W} \subset {\mathcal V}$ a vector subspace. We define 
$$
{\mathcal W}^{\perp,l} = \{v \in {\mathcal V} \; | \; i_{v\wedge w_1
\wedge \cdots \wedge w_l} \Omega = 0,
\hbox{for all} \; w_1, \dots, w_l \in {\mathcal W} \}.
$$

\begin{definition}
${\mathcal W}$ is said to be
\begin{enumerate}
\item $l$-{\rm isotropic} if ${\mathcal W} \subset {\mathcal W}^{\perp,l}$;
\item $l$-{\rm coisotropic} if ${\mathcal W}^{\perp,l} \subset {\mathcal W}$;
\item $l$-{\rm lagrangian} if ${\mathcal W} = {\mathcal W}^{\perp,l}$;
\item {\rm multisymplectic} if ${\mathcal W} \cap {\mathcal W}^{\perp,k} = \{0\}$;
\end{enumerate}
\end{definition}

\begin{proposition}
A subspace ${\mathcal W}$ is $l$-lagrangian if and if it is $l$-isotropic and maximal.
\end{proposition}

\begin{proposition}\label{hala}
Let $V$ an arbitrary vector space. Then:
\begin{enumerate}
\item $V$ is a $k$-lagrangian subspace of ${\mathcal V}_V$ 
and ${\mathcal V}^r_V$, for all $r$;
\item $\Lambda^kV^*$ (resp. $\Lambda^k_rV^*$)
is a 1-isotropic subspace of ${\mathcal V}_V$ (resp. ${\mathcal V}^r_V$).
\end{enumerate}
\end{proposition}

{\bf Proof:}

$(i)$ A direct computation shows that
$$
V^{\perp,k} = \{(x, \gamma) \; | \;
\Omega_V((x, \gamma), (x_1, 0), \dots, (x_k,0)) = 0,
\; \hbox{for all} \; x_1, \dots , x_k \}
$$ 
which is equivalent to the condition $\gamma(x_1, \dots, x_k)=0$
for all $x_1, \dots, x_k \in V$, and therefore
$\gamma = 0$.
Hence $V^{\perp,k}=V$.

The same proof holds for ${\mathcal V}^r_V$.

$(ii)$
We have to prove that
$$
\Lambda^kV^* \subset (\Lambda^kV^*)^{\perp,1}
$$
which is obvious because 
$$
i_{(0,\gamma_1)\wedge (0, \gamma_2)} \, \Omega_V = 0.
$$
The same argument works for ${\mathcal V}^r_V$.
\hfill $\blob$

\begin{remark}{\rm
In addition, notice that
$$
(\Lambda^kV^*)^{\perp,1} = \Lambda^kV^*
$$
which implies that $\Lambda^kV^*$ is in fact 1-lagrangian.
}
\end{remark}

\begin{theorem}\label{teorema1}\cite{M1,M2}
Let $({\mathcal V}, \Omega)$ be a multisymplectic vector space and ${\mathcal W} \subset {\mathcal V}$
a 1-isotropic subspace such that $\dim {\mathcal W} = \dim \Lambda^k({\mathcal V}/{\mathcal W})^*$
and $\dim {\mathcal V}/{\mathcal W} > k$.
Then there exists a $k$-lagrangian subspace $V$
of ${\mathcal V}$ which is transversal to ${\mathcal W}$
(i.e. $V \cap {\mathcal W} = \{0\}$) such that $({\mathcal V}, \Omega)$ is multisymplectomorphic
to the model $({\mathcal V}_V, \Omega_V)$.
\end{theorem}

{\bf Proof:}
\underline{First step:} Define the mapping
\begin{eqnarray*}
\iota: {\mathcal W} & \longrightarrow &\Lambda^k({\mathcal V}/{\mathcal W})^*\\
v & \mapsto & \iota(v) = \widetilde{i_v\Omega}
\end{eqnarray*}
where $\widetilde{i_v\Omega}$ is the induced linear form from
$i_v\Omega \in \Lambda^k {\mathcal V}^*$. Notice that 
$\widetilde{i_v\Omega}$ is well-defined because the isotropic character
of ${\mathcal W}$. In addition, $\iota$ is a linear isomorphism because of
the regularity of $\Omega$.

\underline{Second step:} Such a subspace ${\mathcal W}$ is unique.
First of all, we shall prove that if $u, v \in {\mathcal V}$ are linearly independent
vectors satisfying $i_{u \wedge v} \, \Omega = 0$, then
$\hbox{span} \, (u,v) \cap {\mathcal W} \not= \{0\}$.
Otherwise, we could choose $v_1, \dots, v_{k-2} \in {\mathcal V}$
with $v_i \notin {\mathcal W}$ such that
$\{u, v, v_1, \dots, v_{k-2}\}$ are linearly independent
and $\hbox{span} \, (u,v, v_1, \dots, v_{k-2}) \cap
{\mathcal W} = \{0\}$, because the codimension of
${\mathcal W}$ is at least $k$. But for any $w \in {\mathcal W}$
we would have $i_{w \wedge u \wedge v \wedge v_1 \wedge \cdots
\wedge v_{k-2}} \, \Omega = 0$ which contradicts the fact that
$\iota: {\mathcal W} \longrightarrow \Lambda^k({\mathcal V}/{\mathcal W})^*$
is an isomorphism.

Next, let ${\mathcal W}$ and ${\mathcal W}'$
be two subspaces of ${\mathcal V}$ satisfying the hypothesis of the
theorem. Assume that ${\mathcal W} \not= {\mathcal W}'$; then,
there exists $v \in {\mathcal W}'$ such that $v \notin {\mathcal W}$.
Using the argument above, we deduce that ${\mathcal W}
\cap {\mathcal W}'$ has dimension at least 1.
Consider the subspace $Z = \pi(v) \wedge \Lambda_{k-1} ({\mathcal V}/{\mathcal W})$
of $\Lambda_{k} ({\mathcal V}/{\mathcal W})$, where
$\Lambda_r {\mathcal V}$ is the space of $r$-vectors on ${\mathcal V}$,
and $\pi : {\mathcal V} \longrightarrow {\mathcal V}/{\mathcal W}$ is
the canonical projection.
Of course, $\dim Z > 1$, and we have
$\iota(w)(z)=0$ for any $w \in {\mathcal W} \cap {\mathcal W}'$ and for
any $z \in Z$. Hence we would have 
$w \in \ker \iota$.

\underline{Third step:} There exists a $k$-lagrangian subspace $V$ such that 
${\mathcal V} = {\mathcal W} \oplus V$. Obviously, there are $k$-isotropic
subspaces $U$ such that $U \cap {\mathcal W} = \{0\}$.
To show this last assertion, one could take a vector $v \in {\mathcal V}$
such that $u \notin {\mathcal W}$. It is obvious that
$\hbox{span} \, (u)$ is $k$-isotropic.

Assume that $U \oplus {\mathcal W} = {\mathcal V}$. Then 
${\mathcal W} \cap U^{\perp,k} \subset \ker \iota$ and
hence ${\mathcal W} \cap U^{\perp,k} = \{0\}$.
Therefore $U = U^{\perp, k}$,
and $U$ is $k$-lagrangian.  

Suppose now that $U \oplus {\mathcal W} \neq {\mathcal V}$, then 
$U \neq U^{\perp,k}$; indeed, if
$U = U^{\perp,k}$ (that is, if $U$ were $k$-lagrangian)
then there would be a vector $x \in {\mathcal V}$ such that
$x \notin U \oplus {\mathcal W}$, and then $U \oplus \hbox{span} \, (x)$ would
be $k$-isotropic in contradiction with the maximality of $U$.
Therefore, there is a vector $v \in U^{\perp, k}$ such that
$v \notin U \cup {\mathcal W}$, and we would have a $k$-isotropic
subspace $U' = U \oplus \hbox{span} \, (u)$ such that
$U' \cap {\mathcal W} = \{0\}$. If $U' \oplus {\mathcal W} \not= {\mathcal V}$,
we can repeat the argument and will eventually arrive at a $k$-isotropic
subspace $V$ which is complementary to ${\mathcal W}$.
And using the argument above, we conclude that $V$ is
in fact $k$-lagrangian.

\underline{Fourth step:} Define a linear mapping
\begin{eqnarray*}
\phi & : & {\mathcal W} \longrightarrow \Lambda^kV^*\\
&& \phi(w) = - \frac{1}{k+1} (i_w \Omega)_{|V}
\end{eqnarray*}
A direct computation shows that $\phi$ is an isomorphism.
Next, we define
\begin{eqnarray*}
\psi & : & {\mathcal V} \longrightarrow V \times \Lambda^kV^*\\
&&\psi(v,w) = (v, \phi(w))
\end{eqnarray*}
which is also an isomorphism such that $\psi^* \Omega_V = \Omega$.
\hfill $\blob$

\begin{remark}{\rm
A direct application of Theorem \ref{teorema1} shows that
there exists a basis (a Darboux basis)
$\{e_1, \dots , e_n, f_{\alpha_1 \dots \alpha_k}\}$
such that $\{e_i\}$ is a basis of $V$
and $\{f_{\alpha_1 \dots \alpha_k}\}$ is a basis of ${\mathcal W}$
satisfying the relations
$$
i_{f_{\alpha_1 \dots \alpha_k}} \Omega =
e_{\alpha_1}^* \wedge \cdots \wedge e_{\alpha_k}^*
$$
where $\{e_1^*, \dots, e_n^*\}$ denotes the dual basis of $\{e_1, \dots e_n\}$.
Therefore we have
\begin{equation}\label{cano1}
\Omega = \sum_{\alpha} \, f_{\alpha_1 \dots \alpha_k}^* \wedge
e_{\alpha_1}^* \wedge \cdots \wedge e_{\alpha_k}^*
\end{equation}
where $\{f_{\alpha_1 \dots \alpha_k}^*\}$ is the dual basis
of $\{f_{\alpha_1 \dots \alpha_k}\}$.
}
\end{remark}

\begin{definition}
A triple $({\mathcal V}, \Omega, {\mathcal W})$ satisfying the hypothesis
in Theorem \ref{teorema1} will be called a {\rm multisymplectic vector space 
of type $(k+1, 0)$}.
\end{definition}

\medskip

\begin{theorem}\label{teorema2}
Let $({\mathcal V}, \Omega)$ be a multisymplectic vector space and ${\mathcal W} \subset {\mathcal V}$
a 1-isotropic subspace. Assume that ${\mathcal E} \subset {\mathcal V}/{\mathcal W}$
is a vector subspace of the quotient vector space ${\mathcal V}/{\mathcal W}$.
Let us denote by $\pi : {\mathcal V} \longrightarrow {\mathcal V}/{\mathcal W}$ the canonical projection.
Assume that

\begin{enumerate}
\item $i_{v_1 \wedge \cdots \wedge v_r} \, \Omega = 0$ if $\pi(v_i) \in {\mathcal E}$, for all
$i = 1, \dots , r$;
\item $\dim {\mathcal W} = \dim \Lambda^k_r ({\mathcal V}/{\mathcal W})^*$, where
the horizontal forms are considered with respect to the subspace ${\mathcal E}$;
\item $\dim ({\mathcal V}/{\mathcal W}) > k$.
\end{enumerate}
Then there exists a $k$-lagrangian subspace $V$
of ${\mathcal V}$ which is transversal to ${\mathcal W}$
(i.e., $V \cap {\mathcal W} = \{0\}$) such that $({\mathcal V}, \Omega)$ is multisymplectomorphic
to the model $({\mathcal V}^r_V, \Omega^r_V)$.
\end{theorem}

{\bf Proof:}
First, we define the linear isomorphism
\begin{eqnarray*}
\iota & : & {\mathcal W} \longrightarrow \Lambda^k_r ({\mathcal V}/{\mathcal W})^*\\
&& w \mapsto \iota(w) = \widetilde{i_w \Omega}
\end{eqnarray*}
where $\widetilde{i_w \Omega}$ is the induced $k$-form using that ${\mathcal W}$ is
isotropic and that $\Omega$ satisfies the first condition above.

Next, one follows the arguments given in the
proof of Theorem \ref{teorema1}.
\hfill $\blob$

\begin{remark}{\rm
A direct application of Theorem \ref{teorema2} shows that
the multisymplectic form $\Omega$ can be written as the
canonical multisymplectic form $\Omega_V^r$ on ${\mathcal V}_V^r$
by choosing a convenient Darboux basis.
}
\end{remark}

\begin{definition}
A triple $({\mathcal V}, \Omega, {\mathcal W}, {\mathcal E})$ satisfying the hypothesis
in Theorem \ref{teorema2} will be called a {\rm multisymplectic vector space 
of type $(k+1, r)$}.
\end{definition}

\medskip

Let $({\mathcal V}_1, \Omega_1)$ and  $({\mathcal V}_2, \Omega_2)$ be two multisymplectic vector
spaces of order $k+1$. Take the direct product
${\mathcal V}_1 \times {\mathcal V}_2$ endowed with the $(k+1)$-form
$\Omega_1\ominus \Omega_2 = \pi_1^*\Omega_1 - \pi_2^* \Omega_2$,
where $\pi_1 : {\mathcal V}_1 \times {\mathcal V}_2 \longrightarrow {\mathcal V}_1$
and $\pi_2 : {\mathcal V}_1 \times {\mathcal V}_2 \longrightarrow {\mathcal V}_2$
are the canonical projections. Then
$({\mathcal V}_1 \times {\mathcal V}_2, \Omega_1\ominus \Omega_2)$
is a multisymplectic vector space.

\begin{proposition}\label{lagran1}
Let $({\mathcal V}_1, \Omega_1)$ and  $({\mathcal V}_2, \Omega_2)$ be 
two multisymplectic vector
spaces of order $(k+1)$ and $\phi : {\mathcal V}_1 \longrightarrow 
{\mathcal V}_2$ a linear isomorphism.
Then $\phi$ is a multisymplectomorphism if and only if
its graph is a $k$-lagrangian subspace of the multisymplectic
vector space $({\mathcal V}_1 \times {\mathcal V}_2, \Omega_1\ominus \Omega_2)$.
\end{proposition}

{\bf Proof:}
We recall that
\begin{eqnarray*}
\hskip-1cm(\hbox{graph} \, \phi)^{\perp,k} &=& \{ (x, y) \in {\mathcal V}_1 \times {\mathcal V}_2 \, | \,
(\Omega_1 \ominus \Omega_2)((x,y), (x_1, \phi(x_1)), \dots, (x_k, \phi(x_k)) = 0,\\
&&\forall x_1, \dots, x_k \in {\mathcal V}_1\}
\end{eqnarray*}
Assume that $\phi^* \Omega_2 = \Omega_1$, then if $(x, \phi(x)) \in \hbox{graph} \, \phi$, 
we have 
\begin{eqnarray*}
&&(\Omega_1 \ominus \Omega_2)((x,\phi(x)), (x_1, \phi(x_1)), \dots, (x_k, \phi(x_k))\\
&& = \Omega_1(x, x_1, \dots, x_k) - \Omega_2(\phi(x), \phi(x_1), \dots , \phi(x_k))\\
&&=\Omega_1(x, x_1, \dots, x_k) - \phi^*\Omega_2(x, x_1, \dots , x_k)\\
&& = 0
\end{eqnarray*}
which implies that
$\hbox{graph} \, \phi \subset (\hbox{graph} \, \phi)^{\perp,k}$.
 
Conversely, if $\hbox{graph} \, \phi$ is $k$-isotropic, we have 
$(x, \phi(x)) \in (\hbox{graph} \, \phi)^{\perp,k}$ for all $x \in {\mathcal V}_1$,
and therefore $\phi^* \Omega_2 = \Omega_1$.

In addition, if $\hbox{graph} \, \phi$ is $k$-isotropic, it is also $k$-lagrangian.
In fact, if $(x,y) \in (\hbox{graph} \, \phi)^{\perp,k}$ then we have
$$
\Omega_2(\phi(x)-y, \phi(x_1), \dots, \phi(x_k)) = 0
$$
for all $x_1, \dots , x_k \in {\mathcal V}_1$
and therefore $y=\phi(x)$ because of the regularity of the multisymplectic
form $\Omega_2$ and the fact that $\phi$ is an isomorphism.
\hfill $\blob$

\subsection{Multisymplectic manifolds}

\begin{definition}
A {\rm multisymplectic manifold} $({\mathcal P}, \Omega)$ is a pair consisting of
a manifold ${\mathcal P}$ equipped with a closed $(k+1)$-form $\Omega$ such that
the pair $(T_x{\mathcal P}, \Omega_x)$ is a multisymplectic
vector space for all $x \in {\mathcal P}$. The form $\Omega$ is called
{\rm multisymplectic}.
\end{definition}

\begin{example}{\rm
Let $\Lambda^kM$ be the space of $k$-forms on an arbitrary manifold $M$,
and denote by $\rho : \Lambda^kM \longrightarrow M$ the canonical projection.
We define a canonical $k$-form $\Theta_M^k$ on $\Lambda^kM$ as follows:
$$
\Theta_M^k(\gamma)(X_1, \dots, X_k) = \gamma (T\rho X_1, \dots, T\rho X_k),
$$
for all $X_1, \dots, X_k \in T_\gamma(\Lambda^kM)$ and for all $\gamma \in \Lambda^kM$.

A direct computation shows that $(\Lambda^kM, \Omega_M^k= -d \Theta_M^k)$ 
is a multisymplectic manifold (of order $k+1$).

Assume now that $M$ is a fibred manifold over a manifold $N$, say
$\pi : M \longrightarrow N$ is a fibration. Consider the bundle $\Lambda^k_rM$
of $k$-forms on $M$ which are $r$-horizontal with respect to the fibration
$\pi : M \longrightarrow N$, that is, those $k$-forms $\gamma$ on $M$ such that
$i_{X_1 \wedge \cdots \wedge X_r} \, \gamma = 0$ when 
$X_1, \dots, X_r$ are $\pi$-vertical.
The space $\Lambda^k_rM$ is a submanifold of $\Lambda^kM$, and hence
we have the restriction $(\Theta_M)^k_r$ of $\Theta_M^k$ to $\Lambda^k_rM$.
A simple computation shows that the pair
$(\Lambda^k_rM, (\Omega_M)^k_r = - d(\Theta_M)^k_r)$ is also
a multisymplectic manifold. Of course, we have
$(\Omega_M^k)_{|\Lambda^k_rM} = (\Omega_M)^k_r$.
The canonical projection will be denoted by
$\rho_r : \Lambda^k_r M \longrightarrow M$
t
}
\end{example}

Following the notion of special symplectic manifold introduced by Tulczyjew we can
give the following definition.

\begin{definition}\label{especial}
A {\rm special multisymplectic manifold} $({\mathcal P}, \Omega)$ is a multisymplectic manifold
which is multisymplectomorphic to a bundle of forms. More precisely, $\Omega=-d\Theta$, and 
there exists a diffeomorphism $\alpha : {\mathcal P} \longrightarrow \Lambda^kM$
(or $\alpha : {\mathcal P} \longrightarrow \Lambda^k_rM$),
and a fibration $\pi : {\mathcal P} \longrightarrow M$ such that
$\rho \circ \alpha = \pi$ (resp. $\rho_r \circ \alpha = \pi$)
and $\Theta = \alpha^* \Theta_M^k$
(resp. $\Theta = \alpha^* (\Theta_M)^k_r$).
\end{definition}

\begin{definition}
Let ${\mathcal N}$ be a submanifold of a multisymplectic manifold
$({\mathcal P}, \Omega)$ of order $k+1$. ${\mathcal N}$ is said to be 
$l$-{\rm isotropic} (resp. $l$-{\rm coisotropic}, $l$-{\rm lagrangian}, 
{\rm multisymplectic})
if $T_x{\mathcal N}$ is a
$l$-isotropic (resp. $l$-coisotropic, $l$-lagrangian, multisymplectic)
vector subspace of the multisymplectic vector space
$(T_x{\mathcal P}, \Omega_x)$ for all $x \in {\mathcal N}$.
\end{definition}

\begin{proposition}
$\,$
\begin{enumerate}
\item The fibres of $\rho : \Lambda^kM \longrightarrow M$ 
(and of $\rho_r : \Lambda^k_rM \longrightarrow M$) are 1-isotropic.
\item The image of a $k$-form $\gamma$ on $M$ (resp. a $r$-horizontal
$k$-form) is $k$-lagrangian if and only if $\gamma$ is closed.
\end{enumerate}
\end{proposition}

{\bf Proof:} It follows from Proposition \ref{hala}.
\hfill $\blob$

If $\gamma$ is a ($r$-horizontal) closed $k$-form on $M$, then
$(-d(\Theta_M)^k_r)_{|\hbox{Im} \gamma} = 0$ which implies that
$((\Theta_M)^k_r)_{|\hbox{Im} \gamma}$ is locally closed, say
$$
((\Theta_M)^k_r)_{|\hbox{Im} \gamma} = d \theta ,
$$
and $\theta$ is called a {\it generating $k$-form}.

\medskip

\begin{definition}
A triple $({\mathcal P}, \Omega, {\mathcal W})$, where
${\mathcal W}$ is a 1-isotropic involutive distribution on $({\mathcal P}, \Omega)$
such that the triple 
$(T_x{\mathcal P}, \Omega_x, {\mathcal W}(x))$ is a multisymplectic vector space
of type $(k+1, 0)$, for all $x \in {\mathcal P}$,
will be called a {\rm multisymplectic manifold 
of type $(k+1, 0)$}.
\end{definition}

\begin{theorem}\label{teorema3}\cite{M2}
Let $({\mathcal P}, \Omega)$ be a multisymplectic manifold of type $(k+1,0)$.
Let ${\mathcal L}$ be a $k$-lagrangian
submanifold such that $T{\mathcal L} \cap {\mathcal W}_{|{\mathcal L}} = \{0\}$. Then there exists
a tubular neighbourhood $U$ of ${\mathcal L}$ in ${\mathcal P}$, and a diffeomorphism
$\Phi : U \longrightarrow V=\Phi(U) \subset  \Lambda^k {\mathcal L}$ into
an open neighbourhood $V$ of the zero cross-section in $\Lambda^k {\mathcal L}$
such that $\Phi^* ((\Omega_{\mathcal L}^k)_{|V}) = \Omega_{|U}$, where
$\Omega_{\mathcal L}^k$ is the canonical multisymplectic $(k+1)$-form on
$\Lambda^k{\mathcal L}$.
\end{theorem}

\begin{remark}
{\rm 
Along the paper, the distribution ${\mathcal W}$
and the corresponding vector bundle $\pi_0 : {\mathcal W} \longrightarrow {\mathcal P}$
over ${\mathcal P}$
will be denoted by the same letter.
}
\end{remark}
{\bf Proof:}

First of all, we recall the relative Poincar\'e lemma, 
which will be very useful in what follows.

\begin{lemma}({\bf Relative Poincar\'e lemma})\label{Poincare}
Let $N$ be a submanifold of a differentiable submanifold $M$,
and let $U$ be a tubular neigbourhood of $N$ with bundle map
$\pi_0 : U \longrightarrow N$. Notice that $\pi_0 : U \longrightarrow N$
is a vector bundle. Denote by $\Delta$ the dilation
vector field of this vector bundle, and let $\varphi_t$ be
the multiplication by $t$.
If we define an integral operator on forms on $U$ as follows
$$
I(\Omega)_p = \int_{0}^1 \, i_{\Delta_t} \, \varphi_t^* \Omega_p dt
$$
where $\Delta_t = \frac{1}{t} \Delta$, and $p \in U$, then we have
$$
I(d\Omega) + d(I\Omega) = \Omega - \pi_0^* (\Omega_{|N})
$$
where $\Omega_{|N}$ is the form on $N$ obtained by restricting
$\Omega$ pointwise to $TN$ (observe that $U$ can be taken as a 
normal bundle of $TN$ in $M$).
\end{lemma}

Next, we shall prove the following result.

\begin{lemma}\label{lemauno}
Let $({\mathcal P}, \Omega, {\mathcal W})$ be a multisymplectic
manifold of type $(k+1,0)$.
Let ${\mathcal L}$ be a $k$-lagrangian submanifold of ${\mathcal P}$ which
is complementary to ${\mathcal W}$ (that is, $T{\mathcal L} \oplus {\mathcal W}_{|{\mathcal L}} 
= T{\mathcal P}_{|{\mathcal L}}$).
Then there is a tubular neighbourhood $U$ of ${\mathcal L}$
and a diffeomorphism $\Phi : U \longrightarrow V \subset \Lambda^k {\mathcal L}$
where $V$ is an neighbourhood of the zero section, such that
$\Phi_{|{\mathcal L}}$ is the standard identification of ${\mathcal L}$ with the
zero section of $\Lambda^k{\mathcal L}$, and
$$
\Phi^* ((\Omega^k_{\mathcal L})_{|V}) = \Omega_{|U}.
$$
\end{lemma}

{\bf Proof of Lemma \ref{lemauno}}

Firstly, we define a vector bundle morphism
over the identity of ${\mathcal L}$ by 
$$
\phi (w) = -\frac{1}{k+1} \, i_w \, \Omega.
$$

\begin{figure}[h]
\centering
\setlength{\unitlength}{1cm}
\begin{center}
\begin{picture}(4,2.5)(-0.7,0)
\put(0,2){\makebox(0,0)[r]{${\mathcal W}$}}
\put(4,2){\makebox(0,0)[l]{$\Lambda^k{\mathcal L}$}}
\put(2,0){\makebox(0,0)[c]{${\mathcal L}$}}
\put(0.2,2){\vector(1,0){3.6}}
\put(2,2.3){\makebox(0,0)[r]{$\phi$}}
\put(0.2,1.8){\vector(1,-1){1.6}}
\put(3.8,1.8){\vector(-1,-1){1.6}}
\put(0.8,0.8){\makebox(0,0)[r]{$\pi_0$}}
\put(3.2,0.8){\makebox(0,0)[l]{$\rho$}}
\end{picture}
\end{center}
\centering
\end{figure}
Obviously $\phi$ is injective, and since the dimensionality assumptions,
we deduce that $\phi$ is in fact a vector bundle isomorphism
(see the diagram).

Since $T{\mathcal P}_{|{\mathcal L}} = T{\mathcal L} \oplus {\mathcal W}_{|{\mathcal L}}$, then $\phi$
induces a diffeomorphism on a tubular neighbourhood defined by ${\mathcal W}$
onto a neighbourhood of ${\mathcal L}$ in $\Lambda^k{\mathcal L}$
(as usual, the latter embedding is understood as the identification of
${\mathcal L}$ with the zero section).
We shall denote the restriction of $\phi$ to this tubular neigbourhood by $f$.
Notice that the restriction of $f$ to ${\mathcal L}$ is just the identity,
so that $Tf$ is also the identity on $T{\mathcal L}$; on the other hand,
$Tf$ restricted to ${\mathcal W}$ coincides with $\phi$ because it is 
fiberwise linear. Using the identifications
$T{\mathcal P}_{|{\mathcal L}} = T{\mathcal L} \oplus {\mathcal W}_{|{\mathcal L}}$ and
$T\Lambda^k{\mathcal L}_{|{\mathcal L}} = T{\mathcal L} \oplus \Lambda^k{\mathcal L}$,
we have
\begin{eqnarray*}
f^* \Omega_{\mathcal L}^k ((v_1, w_1), \dots, (v_{k+1}, w_{k+1})) &=&
\Omega_{\mathcal L}^k ((v_1, \phi(w_1), \dots, (v_{k+1}, \phi(w_{k+1}))\\
&=& \, \sum_{i=1}^{k+1} \, (-1)^i \, \phi(w_i)(v_1, \dots, \check{v}_i, \dots, v_{k+1})\\
&=& \, \sum_{i=1}^{k+1} \, \frac{1}{k+1} \, \Omega(v_1, \dots, w_i, \dots, v_{k+1})\\
&=& \, \Omega((v_1, w_1), \dots, (v_{k+1}, w_{k+1}))
\end{eqnarray*}
which implies $f^*\Omega_{\mathcal L}^k = \Omega$ on ${\mathcal L}$.

Next, we use $f$ to pushforward $\Omega$ to obtain a $k+1$-form
$\Omega_1$ in a neighbourhood of ${\mathcal L}$ in $\Lambda^k{\mathcal L}$.
Using Lemma \ref{Poincare} we deduce that
$\Omega_1 = d\Theta_1$, where $\Theta_1 = I(\Omega_1)$.
Recall that $\Omega_{\mathcal L}^k = - d \Theta_{\mathcal L}^k$, and
\begin{equation}\label{aster}
(\Theta_{\mathcal L}^k)_{|{\mathcal L}} = (\Theta_{1})_{|{\mathcal L}} = 0
\end{equation}
because of the definition of $I$.
Define
$$
\Omega_t = \Omega_{\mathcal L}^k + t(\Omega_1 - \Omega_{\mathcal L}^k), \qquad t \in [0,1].
$$
Since
$$
(\Omega_t)_{|{\mathcal L}} = (\Omega_{\mathcal L}^k)_{|{\mathcal L}} = (\Omega_1)_{|{\mathcal L}}
$$
is non-singular, and this is an ``open condition", we
can find a neighbourhood of ${\mathcal L}$ in $\Lambda^k{\mathcal L}$
on which all $\Omega_t$ are non-singular for all $t \in [0,1]$.
In addition, ${\mathcal W}_{\mathcal L} = \ker \{T\rho : T\Lambda^k{\mathcal L} \longrightarrow T{\mathcal L}\}$
is 1-isotropic for all $\Omega_t$, in such a way that
$(\Lambda^k{\mathcal L}, \Omega_t, {\mathcal W}_{\mathcal L})$ is a multisymplectic
manifold of type $(k+1, 0)$, for all $t$.
Notice that
$\Omega_1 - \Omega^k_{\mathcal L} = d (\Theta_1 + \Theta^k_{\mathcal L})$.

>From (\ref{aster}) we deduce that there is a unique time-dependent vector field
$X_t$ taking values in ${\mathcal W}_{\mathcal L}$ (in other words,
$\rho$-vertical) such that
$$
i_{X_t} \, \Omega_t = - \Theta^k_{\mathcal L} + \Theta_1.
$$
Since the vector field $X_t$ vanishes on ${\mathcal L}$,
we can find a neighbourhood of ${\mathcal L}$
in $\Lambda^k{\mathcal L}$ such that the flow $\varphi_t$ of $X_t$ is defined at least
for all $t \leq 1$. Therefore we have
\begin{eqnarray*}
\frac{d}{dt} (\varphi_t^* \Omega_t) &=& \varphi_t^* (L_{X_t} \, \Omega_t)
+ \varphi_t^* (\frac{d\Omega_t}{dt}) \\
&=& \varphi_t^*(di_{X_t} \Omega_t) + \varphi_t^* (\Omega_1 - \Omega^k_{\mathcal L})\\
&=& \varphi_t^*(- d(\Theta_1 - \Theta^k_{\mathcal L}) + \Omega_1 - \Omega^k_{\mathcal L}) = 0.
\end{eqnarray*}
Then we have
$$
\varphi_1^*\Omega_1 = \varphi_0^* \Omega^k_{\mathcal L} = \Omega^k_{\mathcal L}.
$$
But $(X_t)_{|{\mathcal L}} = 0$ implies $(\varphi_t)_{|{\mathcal L}} = id_{|{\mathcal L}}$,
and then 
we deduce that $\varphi_1 \circ f$ gives the desired local diffeomorphism.
\hfill $\blob$

\begin{lemma}\label{lemados}
Let $({\mathcal P}, \Omega, {\mathcal W})$ be a multisymplectic manifold  of type $(k+1,0)$.
Let ${\mathcal L}'$ be a $k$-isotropic submanifold of ${\mathcal P}$ which
is transversal to ${\mathcal W}$ (that is, $T{\mathcal L}' \cap {\mathcal W}_{|{\mathcal L}'} = \{0\}$).
Then there is a $k$-lagrangian submanifold ${\mathcal L}''$ of ${\mathcal P}$
which is complementary to ${\mathcal W}$ and
contains ${\mathcal L}'$.
\end{lemma}

{\bf Proof of Lemma \ref{lemados}}:

Since ${\mathcal L}'$ is transversal to ${\mathcal W}$ we can 
choose a submanifold ${\mathcal L}''$ of $U'$
such that ${\mathcal L}'$ is a deformation retract of ${\mathcal L}''$, and ${\mathcal L}''$ is
complementary to ${\mathcal W}$. As in the theorem above,
since $T{\mathcal P}_{|{\mathcal L}''} = T{\mathcal L}''
\oplus {\mathcal W}_{|{\mathcal L}''}$, then ${\mathcal W}$ induces a 
tubular neighbourhood of ${\mathcal L}''$
in the usual way: $\pi_1 : U' \longrightarrow {\mathcal L}''$.

Next, we apply the relative Poincar\'e lemma to the restricted
form $\Omega$ to this tubular neigborhood. Therefore, there is a $k$-form $\mu$ on $U'$ such that
$$
d\mu = \Omega - \pi_1^*(\Omega_{|{\mathcal L}''})
$$
(indeed, $\mu = I(\Omega)$). 

Now, we can repeat the construction developed in the proof of Lemma \ref{lemauno}
for the $k+1$-form $d\mu$. In fact, the mapping
$\psi : {\mathcal W} \longrightarrow \Lambda^k{\mathcal L}''$ defined by
$\displaystyle{\psi(u) = - \frac{1}{k+1} \, (i_u \, d\mu})$ is a
vector isomorphism, and it induces a local diffeomorphim 
$g : U'' \subset U' \longrightarrow g(U'') \subset \Lambda^k {\mathcal L}''$;
$g$ restrited to ${\mathcal L}''$ is the identity, and $\psi$
on the fibers. Again we can prove
$$
g^* \Omega^k_{{\mathcal L}''} = d\mu
$$
since $(d\mu)_{|{\mathcal L}''} = 0$. 
Proceeding as in the proof of Lemma \ref{lemauno} we can find
a local diffeomorphism $\Psi$ from a tubular neigbourhood $V$ of ${\mathcal L}''$
onto a neighbourhood of the zero section
of $\Lambda^k {\mathcal L}''$ which maps ${\mathcal L}''$ onto the zero section, and
such that
$$
\Psi^*\Omega^k_{{\mathcal L}''} = \Omega
$$
on $V$.

Now, if $j : {\mathcal L}' \longrightarrow {\mathcal L}''$ is the natural
inclusion, we know that $j$ induces an isomorphism in cohomology.
Therefore $j^*(\Omega_{|{\mathcal L}''}) = \Omega_{|{\mathcal L}'}=0$
implies $[\Omega_{|{\mathcal L}''}]_{DR}=0$, and we deduce that
$\Omega_{|{\mathcal L}''} = d\nu$, for some $k$-form $\nu$ on ${\mathcal L}''$.
A direct computation shows now that
$$
{\mathcal L} = \Psi^{-1} \circ (- \nu)({\mathcal L}'')
$$
is a $k$-lagrangian submanifold in $({\mathcal P}, \Omega)$, and in addition 
$T{\mathcal P}_{|{\mathcal L}} = T{\mathcal L}
\oplus {\mathcal W}_{|{\mathcal L}}$. 
\hfill $\blob$

\begin{corollary}\label{coro2}
A multisymplectic manifold $({\mathcal P}, \Omega, {\mathcal W})$ of type $(k+1,0)$
is locally multisymplectomorphic to a canonical multisymplectic manifold
$\Lambda^kM$ for some manifold $M$. Therefore, there are Darboux coordinates around 
each point of ${\mathcal P}$.
\end{corollary}

{\bf Proof}: We only need to choose a point in Lemma \ref{lemados}, and then
apply Theorem \ref{teorema3}.
 \hfill $\blob$

\begin{definition}
Let $({\mathcal P}, \Omega)$ be a multisymplectic manifold of order $k+1$.
Assume that ${\mathcal W}$ is a 1-isotropic foliation of $({\mathcal P}, \Omega)$, and
${\mathcal E}$ is a ``generalised distribution" on ${\mathcal P}$ in the sense
that ${\mathcal E}(x) \subset T_x{\mathcal P}/{\mathcal W}(x)$ is a vector subspace for all 
$x \in {\mathcal P}$. Assume that the quadruple 
$(T_x{\mathcal P}, \Omega_x, {\mathcal W}(x), {\mathcal E}(x))$ is a multisymplectic vector space
of type $(k+1, r)$, for all $x \in {\mathcal P}$. A
quadruple $({\mathcal P}, \Omega, {\mathcal W}, {\mathcal E})$ satisfying
the conditions in Theorem \ref{teorema4} will be called a {\rm multisymplectic manifold 
of type $(k+1, r)$}.
\end{definition}

\begin{theorem}\label{teorema4}
Let $({\mathcal P}, \Omega, {\mathcal W}, {\mathcal E})$ 
be a multisymplectic manifold of type $(k+1,r)$.
Let ${\mathcal L}$ be a $k$-lagrangian
submanifold such that $T{\mathcal L} \cap {\mathcal W}_{\mathcal L} = \{0\}$. Then there exists
a tubular neighbourhod $U$ of ${\mathcal L}$ in ${\mathcal P}$, and a diffeomorphism
$\Phi : U \longrightarrow V=\Phi(U) \subset  \Lambda^k _r{\mathcal L}$ into
an open neighbourhood $V$ of the zero cross-section in $\Lambda^k {\mathcal L}$
such that $\Phi^* (((\Omega_{\mathcal L})^k_r)_{|V}) = \Omega_{|U}$, where
$(\Omega_{\mathcal L})^k_r$ is the canonical multisymplectic $(k+1)$-form on
$\Lambda^k_r{\mathcal L}$.
\end{theorem}

{\bf Proof:} The proof is a consequence of the following two lemmas,
which are proved in a similar way to Lemma \ref{lemauno} and
Lemma \ref{lemados}.

\begin{lemma}\label{lemauno2}
Let $({\mathcal P}, \Omega, {\mathcal W}, {\mathcal E})$ be a multisymplectic
manifold of type $(k+1,r)$.
Let ${\mathcal L}$ be a $k$-lagrangian submanifold of ${\mathcal P}$ which
is complementary to ${\mathcal W}$.
Then there is a tubular neighbourhood $U$ of ${\mathcal L}$
and a diffeomorphism $\Psi : U \longrightarrow V \subset \Lambda^k_r{\mathcal L}$,
where $V$ is an neighbourhood of the zero section, such that
$\Psi_{|{\mathcal L}}$ is the standard identification of ${\mathcal L}$ with the
zero section of $\Lambda^k_r{\mathcal L}$, and
$$
\Psi^* ((\Omega_{\mathcal L})^k_r)_{|V}) = \Omega_{|U}.
$$
\end{lemma}

\begin{lemma}\label{lemados2}
Let $({\mathcal P}, \Omega, {\mathcal W}, {\mathcal E})$ be
a multisymplectic manifold of type $(k+1,r)$.
Let ${\mathcal L}'$ be a $k$-isotropic submanifold of ${\mathcal P}$ which
is transversal to ${\mathcal W}$.
Then there is a $k$-lagrangian submanifold ${\mathcal L}''$ of ${\mathcal P}$
which is complementary to ${\mathcal W}$ and
contains ${\mathcal L}'$.
\end{lemma}

\begin{corollary}\label{coro3}
A multisymplectic manifold $({\mathcal P}, \Omega, {\mathcal W}, {\mathcal E})$ of type $(k+1,r)$
is locally multisymplectomorphic to a canonical multisymplectic manifold
$\Lambda^k_rM$ for some fibration $M \longrightarrow N$. 
Therefore, there are Darboux coordinates around 
each point of ${\mathcal P}$.
\end{corollary}

{\bf Proof}: We only need to choose a point in Lemma \ref{lemados2}, and then
apply Theorem \ref{teorema4}.
 \hfill $\blob$

\section{Lagrangian and hamiltonian settings for classical field theories}

We remit to \cite{BSF,gimmsy1,gimmsy2,LMNRRRS,LMM1,LMM2,LMD,aitor1,PR} for more details.

\subsection{Lagrangian formalism}
 
Let $\pi_{XY}: Y \longrightarrow X$ be a fibred manifold,
where $X$ is an oriented
$n$-dimensional manifold with volume form $\eta$. We choose fibred coordinates
$(x^\mu, y^i)$ on $Y$ such that
$$
\eta = d^nx = dx^1 \wedge \cdots \wedge dx^n, \qquad \pi_{XY} (x^\mu, y^i) = (x^\mu),
$$
where $\mu= 1, \dots, n$, $i= 1, \dots, m$, and $\dim Y = n+m$.
The notation
$$\displaystyle{d^{n-1}x^\mu =  i_{\frac{\partial}{\partial x^\mu}} \, d^nx}$$
will be very useful, since
$dx^\mu \wedge d^{n-1}x^\mu = d^nx$.

Let $\L : Z \longrightarrow \Lambda^nX$ be a lagrangian density, that is,
$\L$ is an $n$-form on $Z$ along the canonical projection $\pi_{XZ} : Z \longrightarrow X$.
Therefore, $\L = L \eta$, where $L : Z \longrightarrow \R$ is a function on $Z$,
and $\eta$ equally denotes the volume form on $X$ and its lifts to the
different bundles over $X$.

One constructs an $n$-form $\Theta_L$ on $Z$ locally given by
$$
\Theta_L = (L- z^i_\mu \frac{\partial L}{\partial z^i_\mu})
d^nx + \frac{\partial L}{\partial z^i_\mu} dy^i \wedge d^{n-1}x^\mu.
$$
The $(n+1)$-form $\Omega_L = - d\Theta_L$ is called the
Poincar\'e-Cartan form.

The de Donder equation is
\begin{equation}\label{dedonder}
i_{\bf h} \, \Omega_L = (n-1) \Omega_L
\end{equation}
where ${\bf h}$ is a connection in the fibred manifold $\pi_{XZ} : Z \longrightarrow X$.

Indeed, if $\sigma$ is a horizontal section of a solution
${\bf h}$ of (\ref{dedonder}) then $\sigma$ is a critical section of
the variational problem determined by $L$.

If $L$ is regular (that is, the hessian matrix
$$
\left(
\frac{\partial^2 L}{\partial z^i_\mu \partial z^j_\nu} 
\right)
$$
is regular) then such a section $\sigma$ is necessarily a 1-jet prolongation, say
$\sigma = j^1\tau$, where $\tau$ is a section of the fibred
manifold $\pi_{XY} : Y \longrightarrow X$.

If ${\bf h}$ is a solution of equation (\ref{dedonder}) and
$$
{\bf h}(\frac{\partial}{\partial x^\mu})
= \frac{\partial}{\partial x^\mu} + y^i_\mu
\frac{\partial}{\partial y^i}
+ z^i_{\nu \mu} \frac{\partial}{\partial z^i_\nu}
$$
then we have
\begin{equation}\label{lagran}
i_{\bf h} \, \Omega_L = (n-1) \Omega_L
\end{equation}
if and only if
\begin{eqnarray}
(y^j_\nu - z^j_\nu) \, \frac{\partial^2 L}{\partial z^i_\mu \partial z^j_\nu} & = & 0 \label{uno}\\
\frac{\partial L}{\partial y^i} - \frac{\partial^2L}{\partial x^\mu \partial z^i_\mu} 
- y^j_\mu \frac{\partial^2L}{\partial y^j \partial z^i_\mu}
- z^j_{\mu \nu} \frac{\partial^2L}{\partial z^j_\mu \partial z^i_\nu}
+ (y^j_\nu - z^j_\nu) \frac{\partial^2L}{\partial y^i \partial z^j_\nu}&=& 0 \label{dos}
\end{eqnarray}

If $L$ is regular, then Eq. (\ref{uno}) implies $y^j_\nu = z^j_\nu$ and
Eq. (\ref{dos}) becomes
\begin{equation}\label{tres}
 \frac{\partial L}{\partial y^i} - \frac{\partial^2L}{\partial x^\mu \partial z^i_\mu} 
- z^j_\mu \frac{\partial^2L}{\partial y^j \partial z^i_\mu}
- z^j_{\mu \nu} \frac{\partial^L}{\partial z^j_\mu \partial z^i_\nu}
 = 0 
\end{equation}

If ${\bf h}$ is flat (that is, the horizontal distribution is integrable)
and $\sigma : X \longrightarrow Z$ is an integral section, then 
$\sigma = j^1(\pi_{YZ} \circ \sigma)$,
and (\ref{tres}) are nothing but the Euler-Lagrange equations for $L$:
\begin{equation}\label{eulerlagrange}
\frac{\partial L}{\partial y^i} - \sum_{\mu=1}^{n} \, \frac{d}{dx^\mu}
\left(\frac{\partial L}{\partial z^i_\mu}\right) = 0.
\end{equation}

\subsection{Hamiltonian formalism}

Denote by $\Lambda^n Y$ the vector bundle over $Y$ of $n$-forms on $Y$,
and by $\Lambda^n_rY$ its vector subbundle consisting of
those $n$-forms on $Y$ which vanish contracted with at least $r$ vertical
arguments.

We have the short exact sequence of vector bundles over $Y$
$$
0 \longrightarrow \Lambda^n_1 Y \longrightarrow \Lambda^n_2 Y 
\longrightarrow Z^*=\Lambda^n_2Y/\Lambda^n _1Y \longrightarrow 0
$$
We choose coordinates as follows:
\begin{eqnarray*}
\Lambda^n_1Y & : & (x^\mu, y^i, p)\\
\Lambda^n_2Y &:& (x^\mu, y^i, p, p^\mu_i)\\
Z^* &:& (x^\mu, y^i, p^\mu_i)
\end{eqnarray*}
since the generic elements in $\Lambda^n_1Y$
(resp. $\Lambda^n_2Y$) have the form
$p \, d^nx$
(resp. $p \, d^nx + p^\mu_i \, dy^i \wedge d^{n-1}x^\mu$).

In order to have a dynamical evolution in the hamiltonian setting one need to choose
a hamiltonian form $h$ on $Z^*$, that is, a section $h : Z^* \longrightarrow \Lambda^n_2Y$
of the canonical fibration
$pr : \Lambda^n_2Y \longrightarrow Z^*$.

The canonical multisymplectic form $(\Omega_Y)^{n}_2$ on $\Lambda^n_2Y$
induces a multisymplectic form (of the same type)
$$
\Omega_h = h^* (\Omega_Y)^n_2.
$$
If $\Theta_h = h^* (\Theta_Y)^n_2$ then $\Omega_h = - d\Theta_h$.

Since
$$
(\Omega_Y)^n_2 = -dp \wedge d^nx - dp^\mu_i \wedge dy^i \wedge d^{n-1}x^\mu
$$
and
$$
h(x^\mu, y^i, p^\mu_i) = (x^\mu, y^i, p=-H(x^\mu, y^i, p^\mu_i), p^\mu_i)
$$
(in other words,
$h = - H d^nx + p^\mu_i dy^i \wedge d^{n-1}x^\mu)$
we obtain
\begin{equation}\label{omegah}
\Omega_h = dH \wedge d^nx - dp^\mu_i \wedge dy^i \wedge d^{n-1}x^\mu
\end{equation}

Consider a connection ${\bf h}^*$ in the fibred manifold
$\pi_{XZ^*} : Z^* \longrightarrow X$, and assume that
$$
{\bf h}^* (\frac{\partial}{\partial x^\mu}) =
\frac{\partial}{\partial x^\mu} + y^i_\mu \frac{\partial}{\partial y^i}
+ p^\nu_{j\mu} \frac{\partial}{\partial p^\nu_j}.
$$
Then 
\begin{equation}\label{hamilton}
i_{{\bf h}^*} \, \Omega_ h = (n-1) \Omega_h
\end{equation}
if and only if
\begin{eqnarray}\label{Hamilton}
y^i_\mu &=&  \frac{\partial H}{\partial p^\mu_i}\\
\sum_\mu \,p^\mu_{i \mu} & = & - \frac{\partial H}{\partial y^i}
\end{eqnarray}

If $\tau : X \longrightarrow Z^*$ is an integral section of ${\bf h}^*$,
and $\tau(x^\mu) = (x^\mu, y^i(x), p^\mu_i)$, then it satisfies
the Hamilton equations
\begin{eqnarray}\label{Hamilton2}
 \frac{\partial y^i}{\partial x^\mu} &=&  \frac{\partial H}{\partial p^\mu_i}\\
 \sum_\mu \, \frac{\partial p^\mu_i}{\partial x^\mu} & = & - \frac{\partial H}{\partial y^i}
\end{eqnarray}

\subsection{The Legendre transformation}

Let $L$ be a lagrangian. We define the extended Legendre transformation
$$
leg_L : Z \longrightarrow \Lambda^n_2Y
$$
by
$$
leg_L (x^\mu, y^i, z^i_\mu) = (x^\mu, y^i, L - z^i_\mu \frac{\partial L}{\partial z^i_\mu},
\frac{\partial L}{\partial z^i_\mu}),
$$
and the Legendre transformation
$$
Leg_L : Z \longrightarrow Z^*
$$
by $Leg_L = pr \circ leg_L$. A direct computation shows that
$L$ is regular if and only if $Leg_L$ is a local diffeomorphism.
$L$ is said to be hyperregular if $Leg_L$ is a global diffeomorphism.
In such case, $h = leg_L \circ Leg_L^{-1}$ is
a hamiltonian form on $Z^*$.

Since the next diagram
\begin{figure}[h]
\centering
\setlength{\unitlength}{1cm}
\begin{center}
\begin{picture}(4,2.5)(-0.7,0)
\put(0,2){\makebox(0,0)[r]{${Z}$}}
\put(4,2){\makebox(0,0)[l]{$Z^*$}}
\put(2,0){\makebox(0,0)[c]{$Y$}}
\put(0.2,2){\vector(1,0){3.6}}
\put(2,2.3){\makebox(0,0)[r]{$Leg_L$}}
\put(0.2,1.8){\vector(1,-1){1.6}}
\put(3.8,1.8){\vector(-1,-1){1.6}}
\put(0.8,0.8){\makebox(0,0)[r]{$\pi_{YZ}$}}
\put(3.2,0.8){\makebox(0,0)[l]{$\pi_{YZ^*}$}}
\end{picture}
\end{center}
\centering
\end{figure}

\noindent
is commutative and $Leg_L^*(\Theta_h) = \Theta_L$, we deduce that Equations
(\ref{lagran}) and (\ref{hamilton}) are equivalent. This means that
the solutions of both equations are related by the Legendre transformation.

\section{The multisymplectomorphism $\tilde{\alpha}$}

Consider the vector bundle $\Lambda^{n+1}_2Z$ with generic
elements of the form
$$
a_i dy^i \wedge d^nx + b^\mu_i dz^i_\mu \wedge d^nx
$$ 
This allows us to introduce local coordinates $(x^\mu, y^i, z^i_\mu, a_i, b^\mu_i)$
in the manifold $\Lambda^{n+1}_2Z$.

On the other hand, we shall denote by $J^1Z^*$ the manifold
of 1-jets of local sections of the fibred manifold
$\pi_{XZ^*} : Z^* \longrightarrow X$. We have a canonical projection
$$
j^1\pi_{YZ^*} : J^1Z^* \longrightarrow Z
$$
Denote by $(x^\mu, y^i, p^\mu_i, y^i_\nu, p^\mu_{i \nu})$ 
the induced coordinates on
$J^1Z^*$ respect to $\pi_{XZ^*} : Z^* \longrightarrow X$,
such that
$$
j^1\pi_{YZ^*}(x^\mu, y^i, p^\mu_i, y^i_\nu, p^\mu_{i \nu})=
(x^\mu, y^i, y^i_\mu).
$$

Define a mapping
$$
\alpha : J^1Z^* \longrightarrow \Lambda^{n+1}_2Z
$$ 
by
$$
\alpha(x^\mu, y^i, p^\mu_i, y^i_\nu, p^\mu_{i \nu}) =
(x^\mu, y^i, y^i_\mu, \sum_\mu p^\mu_{i \mu}, p^\mu_i).
$$

The mapping $\alpha$ is a surjective submersion, or in other words,
$\alpha : J^1Z^* \longrightarrow \Lambda^{n+1}_2Z$ is a fibred
manifold. In order to obtain a diffeomorphism, we need to ``reduce" the manifold
$J^1Z^*$. To do that, we introduce the following equivalence relation:
$$
j^1_x \sigma_1 \equiv j^1_x \sigma_2 
\; \hbox{if and only if they have the same divergence},
$$
which in local coordinates $(x^\mu, y^i, p^\mu_i, y^i_\nu, p^\mu_{i\nu})$
and $(x^\mu, \bar{y}^i, \bar{p}^\mu_i, \bar{y}^i_\nu, \bar{p}^\mu_{i\nu})$
means
$$
\bar{y}^i = y^i, \quad \bar{p}^\mu_i = p^\mu_i, \quad \bar{y}^i_\nu = y^i_\nu, \quad
\sum_{\mu} \bar{p}^\mu_{i\mu} = \sum_{\mu} p^\mu_{i\mu}.
$$

The corresponding quotient manifold will be denoted by $\widetilde{J^1Z^*}$,
and we have a fibration 
$\tilde{pr} : J^1Z^* \longrightarrow \widetilde{J^1Z^*}$.
The induced mapping
$$
\tilde{\alpha} : \widetilde{J^1Z^*} \longrightarrow \Lambda^{n+1}_2Z
$$
is a diffeomorphism, and we have an induced projection 
$$
\widetilde{j^1\pi_{YZ^*}} : \widetilde{J^1Z^*} \longrightarrow Z
$$

Therefore, we can transport the canonical
multisymplectic $(n+2)$-form \\ $(\Omega_Z)^{n+1}_2=-d (\Theta_Z)^{n+1}_2$ on $\Lambda^{n+1}_2Z$
to $\widetilde{J^1Z^*}$ such that
$(\widetilde{J^1Z^*}, \Omega_{\alpha})$ is a multisymplectic manifold,
where $\Omega_{\alpha} = \tilde{\alpha}^*((\Omega_Z)^{n+1}_2)$.

\begin{remark}{\rm
Following the terminology introduced by W.M. Tulczyjew in the symplectic
context, and accordingly to Definition \ref{especial},
we could call $(\widetilde{J^1Z^*}, \Omega_{\alpha})$ a
special multisymplectic manifold, since it is
multisymplectomorphic to a bundle of forms, and the multisymplectic $(n+2)$-form
is $\Omega_{\alpha} = - d \Theta_{\alpha}$ (where 
$\Theta_{\alpha} = \widetilde{\alpha}^*((\Theta_Z)^{n+1}_2)$.
In addition, the following diagram is commutative:
\begin{figure}[h]
\centering
\setlength{\unitlength}{1cm}
\begin{center}
\begin{picture}(4,2.5)(-0.7,0)
\put(0,2){\makebox(0,0)[r]{$\widetilde{J^1Z^*}$}}
\put(4,2){\makebox(0,0)[l]{$\Lambda^{n+1}_2Z$}}
\put(2,0){\makebox(0,0)[c]{$Z$}}
\put(0.2,2){\vector(1,0){3.6}}
\put(2,2.3){\makebox(0,0)[r]{$\tilde{\alpha}$}}
\put(0.2,1.8){\vector(1,-1){1.6}}
\put(3.8,1.8){\vector(-1,-1){1.6}}
\put(0.8,0.8){\makebox(0,0)[r]{$\widetilde{j^1\pi_{YZ^*}}$}}
\put(3.2,0.8){\makebox(0,0)[l]{$\pi_{Z\Lambda^{n+1}_2Z}$}}
\end{picture}
\end{center}
\centering
\end{figure}
}
\end{remark}

\bigskip

Let $\L : Z \longrightarrow \Lambda^nX$ be a lagrangian density,
that is, $\L$ is an $n$-form on $Z$ along the projection 
$\pi_{XZ} : Z \longrightarrow X$.

We put
$$
{\mathcal N}_{\L} =
\{u \in \widetilde{J^1Z^*}  | 
\left( \widetilde{j^1\pi_{XZ^*}}\right)^* (d \L)_u = (\Theta_{\alpha})_u\}
$$

\begin{theorem}
${\mathcal N}_{\L}$ is a $(n+1)$-lagrangian submanifold
of the multisymplectic manifold 
$(\widetilde{J^1Z^*}, \Omega_{\alpha})$.
In addition, the local equations defining ${\mathcal N}_{\L}$ are just
the Euler-Lagrange equations for $L$, where $\L = L \eta$.
\end{theorem}

{\bf Proof:}
>From the definition it follows that
$$
\tilde{\alpha}({\mathcal N}_{\L}) = \hbox{im} \, d \L,
$$
In addition, we have
\begin{eqnarray*}
(\Theta_Z)^{n+1}_2 & = & a_i dy^i \wedge d^nx + b^\mu_i dz^i_\mu \wedge d^nx\\
\alpha^* ((\Theta_Z)^{n+1}_2) & = & p^\mu_{i\mu} dy^i \wedge d^nx +
p^\mu_i dy^i_\mu \wedge d^nx\\
d\L & = & \frac{\partial L}{\partial y^i} dy^i \wedge d^nx + 
\frac{\partial L}{\partial z^i_\mu} dy^i_\mu \wedge d^nx.
\end{eqnarray*}
Since 
$$
(\widetilde{j^1\pi_{XZ^*}})^* (d \L)= \Theta_{\alpha}
$$
if and only if
$$
\tilde{pr}^*(\widetilde{j^1\pi_{XZ^*}}^* (d \L)- \Theta_{\alpha}) = 0
$$
which is in turn equivalent to
$$
(j^1\pi_{XZ^*})^* (d \L)= \alpha^* (\Theta_Z)^n_2,
$$
we deduce that ${\mathcal N}_{\L}$ is locally defined by
\begin{eqnarray}\label{euler}
\sum_\mu \, p^\mu_{i \mu} & = & \frac{\partial L}{\partial y^i}\\
p^\mu_i & = & \frac{\partial L}{\partial z^i_\mu}
\end{eqnarray}
Equations (\ref{euler}) imply that $\tilde{\alpha}({\mathcal N}_{\L}) = \hbox{Im} \, d\L$,
and hence ${\mathcal N}_{\L}$ is a $(n+1)$-lagrangian submanifold of 
$(\widetilde{J^1Z^*}, \Omega_\alpha)$.

Furthermore, we have
$$
\sum_\mu \, p^\mu_{i \mu} = \sum_\mu \,
\frac{\partial}{\partial x^\mu}(\frac{\partial L}{\partial z^i_\mu})
= \frac{\partial L}{\partial y^i}
$$
which are just the Euler-Lagrange equations for $L$.
\hfill $\blob$

\section{The multisymplectomorphism $\tilde{\beta}$}

Recall that there exists a one-to-one
correspondence between connections in the
fibred manifold
$\pi_{XZ^*} : Z^* \longrightarrow X$ and sections of the 1-jet prolongation
$\pi_{Z^* J^1Z^*} : J^1Z^* \longrightarrow Z^*$.
(At a pointwise level we have a one-to-one correspondence
between horizontal subspaces in the fibred manifold
$\pi_{XZ^*} : Z^* \longrightarrow X$ and 1-jets in $J^1Z^*$.)

Define a mapping
$$
\beta : J^1Z^* \longrightarrow \Lambda^{n+1}_2 Z^*
$$
as follows: given a connection ${\bf h}^*$ in the fibred manifold
$\pi_{XZ^*} : Z^* \longrightarrow X$, we take the $(n+1)$-form
$$
\beta({\bf h}^*) = i_{{\bf h}^*} \, \Omega_h - (n-1) \Omega_h.
$$

An arbitrary $(n+1)$-form in $\Lambda^{n+1}_2Z^*$
is written as
$$
A_i dy^i \wedge d^nx + B^i_\mu dp^\mu_i \wedge d^nx
$$
so that we can introduce local coordinates
$(x^\mu, y^i, p^\mu_i, A_i, B^i_\mu)$ on 
$\Lambda^{n+1}_2Z^*$.

If we put
$$
{\bf h}^* (\frac{\partial}{\partial x^\mu}) =
\frac{\partial}{\partial x^\mu} + y^i_\mu \frac{\partial}{\partial y^i}
+ p^\nu_{j\mu} \frac{\partial}{\partial p^\nu_j}
$$
or, equivalently, 
$$
{\bf h}^* (x^\mu, y^i, p^\mu_i) =
(x^\mu, y^i, p^\mu_i, y^i_\mu, p^\nu_{j\mu})
$$
(when ${\bf h}^*$ is considered as a section of $J^1Z^* \longrightarrow Z^*$),
then a straightforward computation shows that
$$
\beta (x^\mu, y^i, p^\mu_i, y^i_\mu, p^\nu_{i\mu}) =
(x^\mu, y^i, p^\mu_i, \sum_\mu \, p^\mu_{i \mu} + \frac{\partial H}{\partial y^i},
- y^i_\mu + \frac{\partial H}{\partial p^\mu_i}).
$$

The mapping $\beta$ is a surjective submersion. Thus, in order to have a diffeomorphism
we consider the induced mapping
$\tilde{\beta} : \widetilde{J^1Z^*} \longrightarrow \Lambda^{n+1}_2 Z^*$.
Therefore we obtain a commutative diagram

\begin{figure}[h]
\centering
\setlength{\unitlength}{1cm}
\begin{center}
\begin{picture}(4,2.5)(-0.7,0)
\put(0,2){\makebox(0,0)[r]{$\widetilde{J^1Z^*}$}}
\put(4,2){\makebox(0,0)[l]{$\Lambda^{n+1}_2Z^*$}}
\put(2,0){\makebox(0,0)[c]{$Z^*$}}
\put(0.2,2){\vector(1,0){3.6}}
\put(2,2.3){\makebox(0,0)[r]{$\tilde{\beta}$}}
\put(0.2,1.8){\vector(1,-1){1.6}}
\put(3.8,1.8){\vector(-1,-1){1.6}}
\put(0.8,0.8){\makebox(0,0)[r]{$\tilde{\rho}$}}
\put(3.2,0.8){\makebox(0,0)[l]{$\pi_{Z^*\Lambda^{n+1}_2Z^*}$}}
\end{picture}
\end{center}
\centering
\end{figure}

where $\tilde{\rho} : \widetilde{J^1Z^*} \longrightarrow Z^*$
is the induced projection from the canonical one
$\rho : J^1Z^* \longrightarrow Z^*$.

Define a $(n+1)$-form $\Theta_{\beta}$ on $\widetilde{J^1Z^*}$
as $\Theta_{\beta} = \tilde{\beta}^*((\Theta_{Z^*})^{n+1}_2)$.
Therefore, the pair $(\widetilde{J^1Z^*}, \Omega_{\beta})$,
$\Omega_\beta = - d\Theta_\beta$, is a multisymplectic manifold of type
$(n+2, 2)$.

\begin{remark}{\rm
It should be noticed that pair $(\widetilde{J^1Z^*}, \Omega_{\beta})$ is a special
multisymplectic manifold.
}
\end{remark}

\begin{theorem}
Let ${\bf h}^*$ be a solution of the de Donder equation.
Then the projection ${\mathcal N}_{h}$ of the image of 
${\bf h}^*$ by $\tilde{pr}$
is a $(n+1)$-lagrangian submanifold
of the multisymplectic manifold 
$(\widetilde{J^1Z^*}, \Omega_{\beta})$.
In addition, the local equations defining ${\mathcal N}_{h}$ are just
the Hamilton equations for $h$.
\end{theorem}

{\bf Proof:}

Since
$$
(\Theta_{Z^*})^{n+1}_2 = A_i dy^i \wedge d^nx + B^i_\mu dp^\mu_i \wedge d^nx
$$
we have
$$
\beta^*((\Theta_{Z^*})^{n+1}_2) = (p^\mu_{i\mu}+ \frac{\partial H}{\partial y^i})
dy^i \wedge  d^nx + (-y^i_\mu + \frac{\partial H}{\partial p^\mu_i}) dp^\mu_i \wedge d^nx.
$$
Therefore, the projection ${\mathcal N}_{h}$ of the image of 
${\bf h}^*$ by $\tilde{pr}$ is just the inverse image of the zero-cross section
of $\Lambda^{n+1}_2Z^*$, and hence it is a $(n+1)$-lagrangian submanifold
of $(\widetilde{J^1Z^*}, \Omega_\beta)$.

The second part of the theorem follows directly from the
preceding discussion.
\hfill $\blob$

\section{Relating $\tilde{\alpha}$ and $\tilde{\beta}$}

The above constructions are collected in the following diagram:

\begin{figure}[h]
\centering
\setlength{\unitlength}{1cm}
\begin{center}
\begin{picture}(4,2.5)(-0.7,0)
\put(-4,2){\makebox(0,0)[r]{$\Lambda^{n+1}_2Z$}}
\put(0.3,2){\makebox(0,0)[r]{$\widetilde{J^1Z^*}$}}
\put(4,2){\makebox(0,0)[l]{$\Lambda^{n+1}_2Z^*$}}
\put(2,0){\makebox(0,0)[c]{$Z^*$}}
\put(-2,0){\makebox(0,0)[c]{$Z$}}

\put(0.4,2){\vector(1,0){3.4}}
\put(-0.6,2){\vector(-1,0){3.4}}

\put(2,2.3){\makebox(0,0)[r]{$\tilde{\beta}$}}
\put(-3,2.3){\makebox(0,0)[l]{$\tilde{\alpha}$}}

\put(0.2,1.7){\vector(1,-1){1.6}}
\put(3.8,1.7){\vector(-1,-1){1.6}}
\put(-3.8,1.7){\vector(1,-1){1.6}}
\put(-0.2,1.7){\vector(-1,-1){1.6}}

\put(0.8,0.8){\makebox(0,0)[r]{$\tilde{\rho}$}}
\put(3.2,0.8){\makebox(0,0)[l]{$\pi_{Z^*\Lambda^{n+1}_2Z^*}$}}
\put(0.3,0.8){\makebox(0,0)[r]{$\widetilde{j^1\pi_{YZ^*}}$}}
\put(-4.5,0.8){\makebox(0,0)[l]{$\pi_{Z\Lambda^{n+1}_2Z}$}}

\end{picture}
\end{center}
\centering
\end{figure}

Since
\begin{eqnarray*}
\tilde{pr}^*(\Theta_\alpha) &=& p^\mu_{i\mu} dy^i \wedge d^nx +
p^\mu_i dy^i_\mu \wedge d^nx\\
\tilde{pr}^*(\Theta_{\beta}) &=& (p^\mu_{i\mu}+ \frac{\partial H}{\partial y^i})
dy^i \wedge  d^nx + (-y^i_\mu + \frac{\partial H}{\partial p^\mu_i}) dp^\mu_i \wedge d^nx
\end{eqnarray*}
we deduce that
\begin{eqnarray*}
\tilde{pr}^*(\Theta_\alpha - \Theta_\beta) &=&
dh - \left( y^i_\mu dp^\mu_i + p^\mu_i dy^i_\mu \right) \wedge d^nx\\
&=& dh - d(p^\mu_i y^i_\mu) \wedge d^nx\\
&=& d\left( h - (p^\mu_i y^i_\mu) \wedge d^nx\right)
\end{eqnarray*}
which implies that
$\Omega_\alpha = \Omega_\beta$.

\begin{theorem}
Let $L$ be a regular lagrangian, and assume that $h = leg_L \circ (Leg_L)^{-1}$.
Then $N_{\L} = N_h$.
\end{theorem}

\bigskip

{\bf Acknowledgment} This work has been partially supported by 
MICYT (Spain) (Grant BFM2001-2272) and Basque Governement.

\end{document}